\documentclass[aps,prd,reprint,groupedaddress,nofootinbib]{revtex4-1}

\usepackage{amsmath}
\usepackage{graphicx}
\usepackage{color}
\usepackage{bm}
\usepackage{comment}
\usepackage{hyperref}

\newcommand{\eps}{\epsilon}

\newcommand{\calR}{{\cal R}}

\newcommand{\EADM}{\mathsf{E}}
\newcommand{\LADM}{\mathsf{L}}
\newcommand{\beq}{\begin{equation}}
\newcommand{\eeq}{\end{equation}}

\definecolor{red  }{rgb}{1,0,0}
\definecolor{blue }{rgb}{0,0,1}
\definecolor{green}{rgb}{0,1,0}

\begin{document}

\title{Self-force as a cosmic censor in the Kerr overspinning problem}

\author{Marta Colleoni}
\author{Leor Barack}
\author{Abhay G. Shah}
\author{Maarten van de Meent}
\affiliation{Mathematical Sciences, University of Southampton, Southampton SO17 1BJ, United Kingdom}

\date{\today}

\begin{abstract}
It is known that a near-extremal Kerr black hole can be spun up beyond its extremal limit by capturing a test particle. Here we show that overspinning is always averted once back-reaction from the particle's own gravity is properly taken into account. We focus on nonspinning, uncharged, massive particles thrown in along the equatorial plane, and work in the first-order self-force approximation (i.e., we include all relevant corrections to the particle's acceleration through linear order in the ratio, assumed small, between the particle's energy and the black hole's mass). Our calculation is a numerical implementation of a recent analysis by two of us [Phys.\ Rev.\  D {\bf 91}, 104024 (2015)], in which a necessary and sufficient ``censorship'' condition was formulated for the capture scenario, involving certain self-force quantities calculated on the one-parameter family of unstable circular geodesics in the extremal limit. The self-force information accounts both for radiative losses and for the finite-mass correction to the critical value of the impact parameter. Here we obtain the required self-force data, and present strong evidence to suggest that captured particles never drive the black hole beyond its extremal limit. We show, however, that, within our first-order self-force approximation, it is possible to reach the extremal limit with a suitable choice of initial orbital parameters.  To rule out such a possibility would require (currently unavailable) information about higher-order self-force corrections. 
\end{abstract}


\maketitle

\section{Introduction}

Consider a Kerr black hole of mass $M$ and dimensionless spin $a/M=1-\eps^2$, with $\eps\ll 1$; and send a particle of rest mass $\mu\ll M$ into the black hole along the equatorial plane. Ignore back-reaction effects on the particle's motion, and assume it follows a geodesic of the Kerr geometry, with conserved specific energy $E$ and angular momentum $L$. Jacobson and Sotiriou  \cite{js} showed that there exists an open domain in the $\{\mu,E,L\}$ space, for which the final configuration is overextremal: 
\begin{equation}\label{OSgeodesics}
a M +\mu L>(M+\mu E)^2. 
\end{equation}
Later work by Barausse {\it et al.}\ \cite{bck1,bck2} demonstrated that, at least for some orbits in that open domain, radiative losses cannot prevent the overspinning. The purpose of our current work is to show that the overspinning scenario is entirely ruled out once all relevant back-reaction effects are taken into account.

The history of the problem goes a while back. In 1974 Wald \cite{wald} proposed the black hole--particle system as a test bed for weak cosmic censorship. He showed (ignoring back-reaction) that an initially extremal Kerr--Newman black hole cannot be made overextremal by capturing a particle, although parameters can be chosen so that the black hole remains extremal. Particles carrying ``dangerous'' amounts of angular momentum, electric charge and/or spin are never captured; rather, they get deflected away by centrifugal, electrostatic and/or spin-spin coupling forces. In 1999 Hubeny \cite{hub} argued that the situation is different if one starts with a {\it nearly} extremal black hole: focusing on the problem of an electric charge in a Reissner-Nordstr\"om geometry, she showed there is an (open) set of captured orbits that overcharge the black hole in that case. However, she also reasoned that back-reaction from the electromagnetic self-force may well prevent suitable particles from being captured. 

Recent work has gone some way towards confirming that expectation. Isoyama, Sago and Tanaka \cite{soich} considered an electric charge released from the point of quasi-static equilibrium---an initial configuration which they represented by the exact (static) ``double Reissner-Nordstr\"{o}m'' solution, whose total Arnowitt--Deser--Misner (ADM) energy and charge are known analytically. Showing that radiative losses from the subsequent plunge into the black hole are negligible, Isoyama {\it et al.}~calculated that the final configuration cannot be that of an overcharged geometry. They further argued that the same conclusion must then apply to any radially falling charge.  
In a later work, Zimmerman, Vega and Poisson \cite{zimm} presented an explicit (numerical) calculation of the charged particle's trajectory including the full effect of the electromagnetic self-force. Analyzing a large sample of orbits within the domain identified by Hubeny, they found no example of successful overcharging: All particles with charge and energy suitable for overcharging the black hole were found to be repelled before reaching the horizon. 

However, both above analyses \cite{soich,zimm} have neglected the coupling to gravity (both the back-reaction from the gravitational perturbation sourced by the perturbed electromagnetic stress-energy, and the perturbation in the electromagnetic field due to the particle's mass), which cannot be easily justified for the problem at hand. A complete analysis would require a calculation of both electromagnetic and gravitational self-forces in the coupled problem, which remains a difficult challenge despite recent progress \cite{zimmerman,tlinz,Zimmerman:2015rga}.

Here we instead consider the purely gravitational problem of an electrically-neutral particle in Kerr geometry. Thus evading the coupling problem comes at the obvious cost of giving up the convenience of working in a spherically symmetric background. However, the current state of affairs in self-force calculations is such that computations on a Kerr background are now routine (see, e.g., the recent \cite{vandeMeent:2015lxa}). Here we revisit the problem from this new vantage point, and provide what we believe to be a first example of a complete, self-consistent analysis of the overspinning problem within the first-order self-force approximation. 

The necessary groundwork for our calculation was laid in a previous paper by two of us \cite{CB}, to be referred to in what follows as CB. Focusing on equatorial capture trajectories, and excluding deeply bound orbits (see below), CB formulated a set of ``censorship conditions'' that must be satisfied in order for the overspinning scenario to be ruled out for all such orbits. The conditions are formulated in terms of a certain one-parameter family of geodesic trajectories, and involve the gravitational self-force (GSF) evaluated along the trajectories in the limit of an extremal geometry. In the current work we numerically evaluate the censorship conditions, and establish, with confidence, that they are indeed satisfied.  

Our results reveal some interesting nuances. To describe them we must first review the results of CB in some more detail. CB first identify (working at leading order in $\eps\ll 1$) the complete region in the parameter-space $\{E,L,\eps,\eta:=\mu/M\}$ for which Eq.\ (\ref{OSgeodesics}) is satisfied, i.e.~overspinning occurs, if the GSF is ignored and the particle follows a geodesic of the Kerr background.  It is found that, for any $E>1$, overspinning occurs in certain narrow ranges ($\propto\eps$) of $\eta,L$ values. Geodesics with $E\leq 1$ cannot overspin. Focusing, therefore, on particles sent in from infinity, CB proceed to incorporate all relevant GSF effects through $O(\eta^2)$ in Eq.\ (\ref{OSgeodesics}). This includes gravitational-radiation losses, as well as the (more subtle but not less important) $O(\eta)$ correction to the critical value of impact parameter for capture. Since the censorship condition is formulated in terms of background (``black-hole centred'') coordinates, which is beneficial in practice for GSF calculations, rather than in a ``center-of-mass'' system where the identification of ADM quantities is easier, care is taken in expressing relevant ADM quantities in terms of background-related quantities through the required order [$O(\eta^2)$ in Eq.\ (\ref{OSgeodesics})]. CB thus obtain an inequality involving the initial energy and angular momentum (at infinity), as well as $\eta$, $\eps$ and the GSF, which describes a condition for the final configuration to remain subextremal. 

CB then show that the above censorship condition is satisfied for all capture trajectories (and for all sufficiently small $\eta,\eps$) if and only if it is satisfied for the one-parameter family of ``critical'' orbits---those lying on the scatter--capture separatrix in the space of initial conditions. Thus, the problem may be conveniently reduced to the question of whether critical orbits satisfy the censorship condition. CB parametrize such orbits by the energy-at-infinity, and formulate a necessary and sufficient censorship condition on that one-parameter family. 

The behavior of critical orbits is subtle when radiation is taken into account. CB distinguish between ``strong'' (exponential) and ``weak'' (power-law) fine-tuning of the initial conditions, depending on the proximity to exact criticality.\footnote{In CB's terminology, these two types are respectively referred to as ``fine-tuned'' and ``generic'' near-critical orbits. We prefer here ``strong'' and ``weak'' fine-tuning, to avoid possible confusion. Both types are fine-tuned, but at a different level.}  {\it Strongly} fine-tuned orbits radiate a significant fraction of their initial energy as they evolve adiabatically through a sequence of unstable circular orbits; in the most extreme case (``perfect'' fine-tuning) ``all'' the orbital energy is radiated away and the particle plunges upon reaching the innermost stable circular orbit (ISCO) \cite{gund}. {\it Weak} fine-tuning still results in trapping on quasi-circular orbits, formally for an infinite amount of time as $\eta\to 0$, but the fraction of energy emitted is vanishingly small in that limit. (A more precise definition of weak and strong fine-tuning will be given in Sec.\ \ref{Sec:review}.) 

CB show that, in the case of weak fine-tuning, all radiation-related terms drop out of the censorship condition through the relevant order [$O(\eta^2)$]. Only one GSF term remains, associated with the conservative correction to the critical value of the angular momentum (at fixed initial energy). CB describe two methods for calculating this correction. The first, direct method involves integration of the GSF along the critical geodesics all the way in from infinity and down to the limiting (unstable) circular orbit. The second method, which relies on a recent perturbative formulation of the conservative Hamiltonian for circular orbits \cite{Isoyama14,hami,Tiec:2015cxa}, requires information about the local metric perturbation (not the GSF itself) only on the limiting circular orbit. The second method is much more easily implemented, and will be used here to compute the shift in the critical value of the angular momentum. 

In the case of strong fine-tuning, radiative terms enter the censorship condition already at the leading order in $\eta$. CB show that the additional information required in this case amounts to a single function of the circular-orbit energy, namely the ratio $\calR(E)$ between the flux of energy absorbed by the black hole and the flux of energy radiated to null infinity.

Thus, two separate censorship conditions emerge. The first applies to weakly fine-tuned orbits and involves only information about the conservative piece of the GSF. It has a remarkably simple form, especially in conjunction with the Hamiltonian formulation of \cite{Isoyama14,hami,Tiec:2015cxa}. The second censorship condition applies to strongly fine-tuned orbits, and requires also knowledge of the radiative flux ratio $\calR(E)$. Both conditions involve quantities that are formally evaluated in the limits $\eps\to 0$ and $\eta\to 0$. Thus, $\eps$ and $\eta$ themselves do not features in the final form of these conditions. The censorship inequalities should be interpreted as necessary and sufficient conditions for overspinning to be averted {\it for sufficiently small $\eps$ and $\eta$}. However, there is no assumption about the relative magnitudes of these two parameters.  

In the current paper we evaluate both censorship conditions. We consider weakly fine-tuned orbits first. We give a strong numerical evidence to suggest that all such orbits {\em precisely saturate} the overspinning condition, i.e.\ they lead to a precisely extremal final geometry, for any value of the initial energy. 
Next we examine whether over-extremality may be achieved through a strong fine-tuning of the initial conditions. This is an intriguing possibility, not least because violation of censorship has been (famously) demonstrated in the past for other carefully fine-tuned configurations \cite{chop}.  However, we find that strong fine-tuning in our problem always acts to drive the system {\em away} from extremality. Capture of strongly fine-tuned orbits results in a subextremal geometry.

A corollary of the above results is that, within our first-order GSF approximation, captured particles generically leave the black hole subextremal, except in the case of weakly fine-tuned critical orbits, which appear to precisely saturate the censorship condition. In the latter case, one must go beyond the first-order GSF approximation to determine whether saturation actually occurs in the physical problem, and theoretically there even remains the possibility of overspinning. One of our main conclusions is that a definitive answer to the overspinning question cannot be reached within the first-order GSF approximation, just as it could not be reached within the geodesic approximation. This is perhaps disappointing, but also intriguing. 



The rest of the paper is structured as follows. In Section \ref{Sec:review} we give a more detailed review of the CB analysis, and present the censorship conditions. In Section \ref{Sec:conservative} we evaluate the condition in the case of weak fine-tuning and conclude that it is saturated, but overspinning never occurs. In Section \ref{Sec:dissipative} we show that strong fine-tuning promotes censorship in our problem. Section \ref{Sec:conclusions} contains a summary and a discussion of our results and their implications.

Throughout this paper we set $G=c=1$ and use the metric signature $(-,+,+,+)$.

\section{Review of CB: the censorship conditions}\label{Sec:review}

\subsection{Basic setup}

The system in question is a gravitationally-bound binary of mass ratio $\eta:=\mu/M\ll 1$. The larger body is a near-extremal Kerr black hole of spin $aM=M^2(1-\eps^2)$, with $\eps\ll 1$.\footnote{Beware the alternative convention $a/M=(1-\eps^2)^{1/2}\simeq 1-\eps^2/2$ is common in related literature.} The smaller body (``particle'') is a compact object of negligible spin, which, for concreteness and simplicity, we may think of as a small non-spinning black hole. 
We work in black-hole perturbation theory, which is based on a formal expansion in $\eta$ about the background geometry of the large black hole. Specifically, we will consider the binary dynamics in the first-order GSF approximation, namely through the first order in $\eta$ beyond the limit of geodesic motion. We let $\{t,r,\theta,\varphi\}$ denote standard Boyer-Lindquist coordinates on the background geometry, and let the particle be sent in along the equatorial plane, $\theta=\pi/2$. From symmetry it is clear that the orbit will remain equatorial even under the effect of the GSF. 

It is assumed that the standard GSF formulation of the binary dynamics applies: the particle's motion is described locally in terms of an effective (accelerated) trajectory on the background spacetime, which, at our order of approximation, is insensitive to the particle's internal structure. In particular, the particle will be assumed to have fallen into the large black hole if and only if its trajectory had crossed the latter's event horizon. 

It will be useful to define, for given $\{M,\mu,\eps\}$,
\begin{eqnarray}\label{ELADM}
\EADM&:=&(M_{\rm ADM}-M)/\mu,
\\
\LADM&:=&(J_{\rm ADM}-aM)/(\mu M),
\end{eqnarray}
where $M_{\rm ADM}$ and $J_{\rm ADM}$ are the total (conserved) ADM mass and angular momentum of the binary's spacetime. $\EADM$ may be interpreted as the particle's contribution (per $\mu$) to the ADM energy in the ``black hole frame'', and $\LADM$ is the particle's specific (and a-dimensionalized) orbital angular momentum, again in the ``black hole frame''.\footnote{We use $\EADM$ and $\LADM$ in place of CB's $E_{\rm ADM}^{\rm p}$ and $L_{\rm ADM}^{\rm p}$, for notational simplicity.} For the GSF formulation to make sense, we require, in addition to $\eta\ll 1$, also $\EADM\ll 1/\eta$. We do not consider ultrarelativistic particles with $\EADM\gg 1$ (not even ones with $\EADM\ll 1/\eta$), because for such orbits there does not yet exist a rigorous GSF formulation. Nor do we consider the family of ``deeply bound'' orbits discussed in Sec.\ II.B of CB: low-energy orbits that are confined to the immediate exterior of the large hole, below the ISCO radius. CB explain that, for $\eps\ll 1$ and relevant values of $\eta$, such orbits plunge into the black hole within an amount of proper time comparable to the particle's own ``light-crossing'' time. Ignoring the particle's finite extent does not seem justified in that situation, so such orbits require a separate treatment. For simplicity, CB opted to exclude deeply bound orbits from the analysis. In practice, this amounts to assuming that the particle is thrown in from outside the ISCO radius.
  
Our system is thus parametrized by the quartet of values $\{\eta,\eps,\EADM,\LADM\}$, up to an initial orbital phase and up to a trivial overall scale set by $M$, neither of which are of relevance in our analysis. This parameter space splits into two precisely complementary subspaces, one corresponding to orbits that end up crossing to horizon, and another corresponding to orbits that scatter to infinity without crossing the horizon. For captured orbits, let $M_{\rm B}$ and $J_{\rm B}$ be the Bondi energy and angular momentum of the spacetime at the (retarded) time of horizon crossing.  
The question of overspinning then takes the following simple form: Among all configurations $\{\eta,\eps,\EADM,\LADM\}$, with arbitrarily small $\{\eta,\eps\}$, are there any in which the particle crosses the horizon and $J_{\rm B}>M_{\rm B}^2$? If so, then the likely scenario is that a naked singularity is exposed \cite{wald}. If not, then the conjecture of cosmic censorship holds. 


\subsection{Test-particle limit}

Let us first review the situation when back-reaction is ignored and the particle is treated as a test mass (this is the case first analyzed in \cite{js}, but the details below follow CB). The particle moves on a timelike geodesic of the Kerr background metric $g_{\alpha\beta}$, with conserved (specific) energy and angular momentum given by $\EADM=-u_t$ and $\LADM=u_\varphi/M$, where $u^{\alpha}$ is the particle's four-velocity and $u_{\alpha}=g_{\alpha\beta}u^{\beta}$. Assuming the particle is thrown in from outside the ISCO radius, it will cross the horizon if and only if $\LADM$ is smaller than a certain critical value $\LADM_{c}^{(0)}(\EADM;\eps)$, which corresponds to the angular momentum of an unstable circular geodesic orbit with energy $\EADM$ (``peak of the effective potential''). We use a superscript $(0)$ to denote the test-particle limit. CB find, in the near-extremal case, 
\begin{equation}\label{Lc0}
\LADM_c^{(0)}(\EADM;\eps)=2\EADM+(6\EADM^2-2)^{1/2}\eps+O(\eps^2).
\end{equation} 

The Bondi mass and angular momentum are constant and given by $M_{\rm B}=M+\mu \EADM$ and $J_{\rm B}=aM+\mu M \LADM$, respectively. Captured orbits overspin the black hole if and only if they satisfy $J_{\rm B}>M_{\rm B}^2$, which may be rearranged to give $\LADM>2\EADM+\eps^2/\eta+\eta \EADM^2$. Thus, for given $\{\eta,\eps,\EADM\}$, the capture condition bounds $\LADM$ from above, $\LADM<\LADM_{c}^{(0)}$, while the overspinning condition bounds it from below. Neglecting high-order terms in $\eps$, overspinning would occur if the double inequality
\begin{equation}\label{double_ineq}
\eps^2/\eta+\eta \EADM^2<\LADM-2\EADM<(6\EADM^2-2)^{1/2}\eps
\end{equation}
can be satisfied with $\eps,\eta\ll 1$. 

A simple analysis \cite{CB} shows that (\ref{double_ineq}) is never satisfied for $\EADM\leq 1$, but it can be satisfied for any $\EADM>1$ by choosing $\eta$ from within the range  
\begin{equation}\label{mrange}
\eps f_-(\EADM) <\eta <  \eps f_+(\EADM),
\end{equation}
with
$f_{\pm}=\frac{1}{\sqrt{2}\, \EADM^2}\left[\sqrt{3\EADM^2-1}\pm \sqrt{\EADM^2-1}\right]$.
For any such choice, there is an open range of $\LADM$ values satisfying (\ref{double_ineq}).
Thus, for a given $\eps\ll 1$, there is an open overspinning domain in the space of $\{\eta,\EADM, \LADM\}$, described by $\EADM>1$ together with Eqs.\ (\ref{mrange}) and (\ref{double_ineq}). This domain is depicted in Figures 2 and 3 of CB. The key point: if back-reaction effects were negligible, it would be possible to overspin the black hole by throwing a particle in from infinity.  

Equation (\ref{mrange}) shows that, as might be expected, $\eta$ and $\eps$ are of the same order of magnitude in overspinning configurations. Equation (\ref{double_ineq}) shows, in turn, that all overspinning orbits have $\LADM-2\EADM=O(\eta)$. Such orbits are located close to the parameter-space separatrix between captured and scattered orbits. They each exercise $O(\ln\eta)\gg 1$ near-circular revolutions near the peak of the effective potential before plunging into the black hole. The amount of excess spin produced by such orbits is $J_B-M_B^2=O(\eta^2)$. 

This implies that back-reaction effects may qualitatively change the outcome of the analysis. A priori, $O(\eta)$ self-acceleration may shift the value of the critical angular momentum (for a given energy) by an amount of $O(\eta)$, comparable to the width of the overspinning window. Furthermore, radiation of energy and angular momentum in gravitational waves to infinity may, a priori, alter the final excess spin by an amount of $O(\eta^2\ln\eta)$, {\em greater} than the excess spin predicted with radiation neglected.  Clearly, {\em the point-particle approximation is inadequate} in our problem. It is a case where a leading-order perturbative treatment predicts its own inadequateness. By the end of our analysis it will become clear that, remarkably, the same conclusion carries over to the next perturbative order.


\subsection{Self-force approximation and the correction to the capture--scatter separatrix}

In the first-order GSF approximation, the particle's motion is described locally by an accelerated worldline in the Kerr background metric $g_{\alpha\beta}$. The tangent four-velocity, normalized using $g_{\alpha\beta}u^{\alpha}u^{\beta}=-1$, satisfies  $\mu u^{\beta}\nabla_{\beta}u^{\alpha}=F^{\alpha}$, where $\nabla_{\beta}$ is a covariant derivative compatible with $g_{\alpha\beta}$, and $F^{\alpha}(\propto \eta^2)$ is the GSF. The latter may be attributed to a certain smooth (locally-defined) self-potential denoted $h^R_{\alpha\beta}$---the ``R field''---which is a particular solution of the source-free linearized Einstein's equations \cite{DW}. The particle's trajectory may also be interpreted as a geodesic in $g_{\alpha\beta}+h^R_{\alpha\beta}$.

For given $\eps,\eta$, our equatorial orbits may again be parametrized by the ADM-related constants $\EADM,\LADM$ defined in Eq.\ (\ref{ELADM}). These now also carry information about the self-gravity of the particle and about the radiation content of spacetime. For given $\eps,\eta,\EADM$ there exists a critical value $\LADM_c(\EADM;\eps,\eta)$ representing a threshold for immediate capture: orbits with $\LADM>\LADM_c$ scatter off the black hole (at least at first approach; they may still eventually fall into the black hole following subsequent encounters), while those with $\LADM<\LADM_c$ are immediately captured. The geodesic limit, $\eta\to 0$, of $\LADM_c(\EADM;\eps,\eta)$ is $\LADM_c^{(0)}(\EADM;\eps)$, given in Eq.\ (\ref{Lc0}). Not surprisingly, much of the physics relevant to our problem is played out near the capture--scatter threshold. 

In the geodesic case (and for a fixed $\eps$), $L=\LADM_c^{(0)}(\EADM;\eps)$ describes a 1-parameter family of disjoint {\it critical orbits}. Each critical orbit is homoclinic in nature, approaching an (unstable) circular geodesic at $t\to\infty$. The situation becomes more subtle when GSF effects are included, due to dissipation.    For given $\eps,\eta$, the function $L=\LADM_c(\EADM;\eps)$ still describes a 1-parameter family of critical orbits, but these are no longer disjoint. Rather, they all merge to form a ``global attractor''. The attractor may be thought of as a smooth sequence of quasi-circular unstable orbits starting at the ``light ring'' and ending at the ISCO. A critical orbit meets the attractor at a point that depends on its initial energy, and proceeds to evolve radiatively along it until it transits to a plunge near the ISCO. A small perturbation away from the critical value $\LADM=\LADM_c(\EADM;\eps,\eta)$ results in the particle's leaving the attractor before the ISCO is reached. The point of departure may be controlled by fine-tuning the magnitude of $L$ around its critical value.  

When considering near-critical orbits in the overspinning analysis, it is important to distinguish between two cases, differing by how well $\LADM$ is tuned to its critical value. ``Strong'' fine-tuning is one in which $\LADM-\LADM_c\sim \exp(-\alpha/\eta)$, with some positive constant $\alpha$. Such orbits radiate away amounts of $O(\eta)$ energy and angular momentum as they evolve along the attractor [leading to $O(1)$ changes in the values of the particle's specific  energy and angular momentum]. ``Weak'' fine-tuning is one in which $\LADM-\LADM_c\sim \eta^\beta$, with some $\beta\geq 1$. Such orbits radiate away only $O(\eta^2\ln\eta)$ of energy and angular momentum before leaving the attractor. As will be discussed below, CB found that radiation effects enter the overspinning condition at relevant order only for strongly fine-tuned near-critical orbits. For all other orbits, radiation terms are negligible within the first-order GSF approximation.  

The overspinning analysis requires an explicit expression for $\LADM_c(\EADM;\eps,\eta)$ through $O(\eta,\eps)$. For orbits that come from infinity (i.e., ones with $\EADM\geq 1$) CB obtain, to the required order,
\begin{equation}\label{Lc}
\LADM_c(\EADM;\eps,\eta)=\LADM_c^{(0)}(\EADM;\eps)+\delta\LADM_c(\EADM;\eta),
\end{equation}
where $\LADM_c^{(0)}$ is given in Eq.\ (\ref{Lc0}), and the GSF correction is
\begin{equation}\label{dLc}
\delta\LADM_c=-\eta(\EADM^2+1)+\lim_{\eps\to 0}\int_{R_\eps}^{\infty}\left(2\tilde F_t+\tilde F_{\varphi}\right)dr/u^r.
\end{equation}
Here $\tilde F_t:=F_t/\mu$ and $\tilde F_\varphi:=F_\varphi/(M\mu)$ are components of the self-acceleration ($\propto \eta$), and the integration is carried out along the critical {\em geodesic} with energy $\EADM$ and angular momentum $\LADM_c^{(0)}(\EADM;\eps)$. The cutoff radius $r=R_{\eps}(\EADM;\eps)$ is that of the corresponding unstable circular geodesic. The term $-\eta(\EADM^2+1)$ arises from the transformation between the background-defined energy-at-infinity and angular-momentum-at-infinity, $-u_{t}(r\to\infty)$ and $u_{\varphi}(r\to\infty)$ respectively (these are used in the GSF description of the motion), and the ADM-related quantities $\EADM$ and $\LADM$---see Eq. (72) and Appendix A of CB.


\subsection{Censorship conditions}

Consider an orbit that ends up crossing the event horizon, and let ${\cal E}^+$ and ${\cal L}^+$ be the total energy and angular momentum in gravitational waves radiated out to null infinity up until the retarded time of horizon crossing. The Bondi mass and angular momentum corresponding to that retarded time are $J_B=M+\mu \EADM-{\cal E}^+$ and $aM +\mu \LADM-{\cal L}^+$, respectively. The condition for overspinning is $J_B>M_B^2$, which, upon substituting $a/M=1-\eps^2$ and rearranging, reads
\begin{equation} \label{OS1}
\eps^2+\eta(2\EADM-\LADM)-{\cal W}^+ + (\eta\EADM-{\cal E}^+)^2<0,
\end{equation}
where ${\cal W}^+:=2{\cal E}^+-{\cal L}^+$.
CB showed that the left-hand side of this inequality is minimized (with respect to $\LADM$, at fixed $\EADM,\eps,\eta$) by near-critical orbits with $\LADM=\LADM_c(\EADM;\eps,\eta)+o(\eta)$. Thus it suffices to restrict attention to this type of orbits. If, for some $\EADM$ (and $\eps,\eta\ll 1$), the inequality (\ref{OS1}) can be shown to apply for a near-critical orbit, that would establish a violation of censorship. If, conversely, (\ref{OS1}) can be shown not to be satisfied even for near-critical orbits (for any $\EADM$), then it cannot be satisfied for any captured orbit, and censorship is protected. We can therefore proceed by substituting  $\LADM=\LADM_c(\EADM;\eps,\eta)$ in Eq.\ (\ref{OS1}). Using (\ref{Lc}) and (\ref{Lc0}) one thus arrives at the {\it censorship} condition
\begin{equation} \label{OS2}
\eps^2 + \eps\eta \phi(\EADM)-\eta\delta\LADM_c(\EADM)-{\cal W}^+ +(\eta\EADM-{\cal E}^+)^2\geq 0,
\end{equation}
where 
\begin{equation}\label{phi}
\phi(\EADM):=-(6\EADM^2-2)^{1/2}.
\end{equation} 
Censorship is protected if and only if (\ref{OS2}) is satisfied for all $\EADM$ when $\eps,\eta$ are arbitrarily small.

CB next showed that (\ref{OS2}) can be written in the convenient equivalent form 
\begin{equation} \label{OS3}
\eps^2+ \eps\eta \phi(\EADM)-\eta\delta\LADM_c^{\rm cons}(\EADM)-{\cal W}^+_{\rm (qc)} +(\eta\EADM-{\cal E}^+_{\rm (qc)})^2\geq 0,
\end{equation}
where $\delta\LADM_c^{\rm cons}$ is the contribution to $\delta\LADM_c$ from the {\it conservative} piece of the GSF, and ${\cal W}^+_{\rm (qc)},{\cal E}^+_{\rm (qc)}$ are the contributions to ${\cal W}^+,{\cal E}^+$ from the quasi-circular part of the near-critical orbit. [CB established this by showing that the dissipative contribution to $\delta\LADM_c$ cancels against the part of ${\cal W}^+$ corresponding to the ``approach'' part of the critical orbit; that the contribution to ${\cal E}^+$ from the approach is negligible in Eq.\ (\ref{OS2}); and that the contributions to both ${\cal W}^+$ and ${\cal E}^+$ from the final plunge from the attractor to the horizon are also negligible in that equation.] In Eq.\ (\ref{OS3}), $\delta\LADM_c^{\rm cons}$ is an $O(\eta)$ quantity for any near-critical orbit, while the $\eta$-scaling of the radiative terms ${\cal W}^+_{\rm (qc)}$ and ${\cal E}^+_{\rm (qc)}$ depends on the degree of fine-tuning. For weakly fine-tuned orbits CB find ${\cal W}^+_{\rm (qc)}=O(\eps)O(\eta^2\ln\eta)$ and ${\cal E}^+_{\rm (qc)}=O(\eta^2\ln\eta)$, while for strong fine-tuning the scaling is instead ${\cal W}^+_{\rm (qc)}=O(\eps)O(\eta)$ and ${\cal E}^+_{\rm (qc)}=O(\eta)$. 

It follows that, for weak fine-tuning, both radiative terms ${\cal W}^+_{\rm (qc)}$ and ${\cal E}^+_{\rm (qc)}$ are subdominant in Eq.\ (\ref{OS3}) and drop out at our working order (we assume here $\eps|\ln\eta|\ll 1$). The two terms survive in Eq.\ (\ref{OS3}) only for strongly fine-tuned orbits. We proceed by considering the two cases separately. 

\subsubsection{Weak fine-tuning} 

In the case of weak fine-tuning we are left with the simple censorship condition 
\begin{equation} \label{OS4}
\eps^2+\eta\eps \phi(\EADM)+\eta^2 \psi(\EADM)\geq 0,
\end{equation}
with
$\psi(\EADM):=\EADM^2-\delta\breve\LADM_c^{\rm cons}(\EADM)$. We have introduced 
\begin{equation} \label{brevedL}
\delta\breve\LADM_c^{\rm cons}:=\eta^{-1}\delta\LADM_c^{\rm cons},
\end{equation}
so that both coefficients $\phi(\EADM)$ and $\psi(\EADM)$ in Eq.\ (\ref{OS4}) have finite (generally nonzero) limits $\eta\to 0$ and $\eps\to 0$ (taken with a fixed $\EADM$). Censorship holds if and only if, for any $\EADM$, the inequality in Eq.\ (\ref{OS4}) is satisfied for all (small, positive) $\eta,\eps$. A simple analysis shows \cite{CB} that this demands $\psi\geq \phi^2/4$, leading to the necessary and sufficient censorship condition
\begin{equation} \label{OS_weak}
\delta\breve\LADM_c^{\rm cons}(\EADM)\leq \frac{1}{2}(1-\EADM^2) ,
\end{equation}
for all $\EADM$. The condition is {\it sufficient} in the sense that its confirmation would establish that overspinning is not possible for any $\eta,\eps\ll 1$ and any $\EADM$ (when strongly fine-tuned orbits are excluded). The condition (\ref{OS_weak}) is {\it necessary} in the sense that its violation for some $\EADM$ would imply that, for that value of $\EADM$, there exist $\eta,\eps\ll 1$ with which overspinning can be achieved. 

In Eq.\ (\ref{OS_weak}) we may allow $\EADM$ to vary in the full range $\EADM_{\rm isco}<\EADM\ll 1/\eta$, where $\EADM_{\rm isco}=\frac{1}{\sqrt{3}}$ is the ISCO value of $\EADM$ in the limit $\eta,\eps\to 0$. This is notwithstanding the fact that the explicit expression given for $\delta\LADM_c^{\rm cons}$ in Eq.\ (\ref{dLc}) (with the full GSF $\tilde F_{\alpha}$ replaced with its conservative piece) only applies for $\EADM\geq 1$. In subsection \ref{subsec:z} below we will derive an alternative expression for $\delta\LADM_c^{\rm cons}$, applicable for any $\EADM$ in the above full range. 

\subsubsection{Strong fine-tuning} 

First, however, let us formulate a censorship condition for strongly fine-tuned orbits. As mentioned, in that case one has ${\cal W}^+_{\rm (qc)}=O(\eps)O(\eta)$ and ${\cal E}^+_{\rm (qc)}=O(\eta)$ and both terms feature already at leading order in Eq.\ (\ref{OS3}). We introduce the rescaled quantities 
\begin{equation}
\breve{\cal W}^+_{\rm (qc)}:=(\eta\eps)^{-1}{\cal W}^+_{\rm (qc)},
\quad\quad
\breve{\cal E}^+_{\rm (qc)}:=\eta^{-1}{\cal E}^+_{\rm (qc)},
\end{equation}
which should have finite (generally nonzero) limits $\eta,\eps\to 0$. We note that $\breve{\cal W}^+_{\rm (qc)}$ and $\breve{\cal E}^+_{\rm (qc)}$ depend not only on $\EADM$ but also on the precise fine-tuning. It is then convenient to re-parametrize the problem using the pair $\{E_i,E_f\}:=\{\EADM,\EADM-\cal{E}/\eta\}$, where $\cal{E}$ is the total energy radiated away (both to infinity and down the black hole) during the quasi-circular whirl. To leading order in $\eta$, $E_i$ is the particle's specific energy just upon entering the whirl, and $E_f$ is its specific energy just before leaving it. The difference $E_i-E_f$ is the total energy (per $\eta$) radiated during the whirl. Any value of $E_f$ in the range $\EADM_{\rm isco}< E_f<E_i$ may be obtained via strong fine-tuning.  

The censorship condition (\ref{OS3}) now takes the form 
\begin{equation} \label{OS5}
\eps^2+\eta\eps \tilde\phi(E_i,E_f)+\eta^2 \tilde\psi(E_i,E_f)\geq 0,
\end{equation}
with
\begin{eqnarray} \label{tildephi}
\tilde\phi&=&-(6E_i^2-2)^{1/2}-\breve{\cal W}^+_{\rm (qc)}, \\  \label{tildepsi}
\tilde\psi&=&-\delta\breve\LADM_c^{\rm cons}(E_i)+(E_i-\breve{\cal E}^+_{\rm (qc)})^2.
\end{eqnarray} 
Here we have used the fact that $\EADM=E_i$ to leading order in $\eta$. $\tilde\phi$ and $\tilde\psi$ depend on $E_f$ through $\breve{\cal W}^+_{\rm (qc)}$ and $\breve{\cal E}^+_{\rm (qc)}$, respectively. Censorship holds if and only if, for any $E_i,E_f$ satisfying $\EADM_{\rm isco}< E_f<E_i$, the inequality in Eq.\ (\ref{OS5}) is satisfied for all (small, positive) $\eta,\eps$. A simple analysis shows \cite{CB} that this happens if and only if
\begin{equation}\label{OS_strong}
\tilde\psi \geq \left(\min\{\tilde\phi/2,0\}\right)^2
\end{equation}
for any $\EADM_{\rm isco}< E_f<E_i$.
This constitutes a necessary and sufficient censorship condition for strongly fine-tuned orbits.

The evaluation of (\ref{OS_strong}) requires, in addition to $\delta\breve\LADM_c^{\rm cons}$, also the radiative quantities $\breve{\cal E}^+_{\rm (qc)}$ and $\breve{\cal W}^+_{\rm (qc)}$. CB show that, at the required order, these can be conveniently obtained using 
\begin{equation}\label{calE+}
\breve{\cal E}^+_{\rm (qc)}(E_i,E_f)=\int_{E_f}^{E_i}\frac{dE}{1+{\cal R}(E)},
\end{equation}
\begin{equation}\label{calW+}
\breve{\cal W}^+_{\rm (qc)}(E_i,E_f)=-\int_{E_f}^{E_i}\frac{b(E)}{1+{\cal R}(E)}\,dE,
\end{equation}
where 
\begin{equation}\label{b}
b(E):=6E(6E^2-2)^{-1/2}.
\end{equation}
Here ${\cal R}(E)$ is the ratio between the flux of energy absorbed by the black hole and the flux of energy radiated to infinity, for a particle on a circular geodesic orbit with specific energy $E$, in the extremal limit $\eps\to 0$. Hence, the only information we require about radiation is encapsulated in a single dimensionless function ${\cal R}(E)$, evaluated on circular geodesics. We note ${\cal R}<0$ for $E<\frac{2}{\sqrt{3}}$, the superradiant regime (we will show how this result is arrived at in Sec.\ \ref{Sec:dissipative}). However, we also have ${\cal R}>-1$ for any $E$, implied by the known non-existence of ``floating'' orbits. Since the integrand in (\ref{calW+}) is positive definite (and $E_i>E_f$), we have $\breve{\cal W}^+_{\rm (qc)}<0$, so the sign of $\tilde\phi$ in Eq.\ (\ref{OS_strong}) is not a priori known.

\subsection{Reexpressing $\delta\LADM_c^{\rm cons}$ in terms of redshift}\label{subsec:z}

Calculating $\delta\LADM_c^{\rm cons}$ using Eq.\ (\ref{dLc}) has two drawbacks. First, this formula applies only to particles thrown in from infinity (i.e., ones with $\EADM\geq 1$). This restriction comes from the fact that in deriving (\ref{dLc}) CB relied on having at hand an explicit relation, correct through $O(\eta^2)$, between the ADM mass and angular momentum on one hand, and background-defined quantities like the four-velocity $u^{\alpha}$ on the other hand. Such a relation could only be derived under the condition that the initial binary separation was infinitely large.  Although the test-particle analysis provides some motivation for concentrating on such orbits, it is preferable to relax the restriction $\EADM\geq 1$ when considering the GSF case. 

The second drawback of the formula (\ref{dLc}) is a practical one. Implementing the formula requires a calculation of the GSF along unbound orbits. However, existing GSF calculation methods and working codes are specialized to {\em bound}, quasi-periodic orbits. Adapting these codes to deal with unbound orbits might be possible in principle, but would require much development of new method and code.  

CB derived an alternative formula for $\delta\LADM_c^{\rm cons}$, circumventing both problems. Their derivation built on recent work by Isoyama, Le Tiec and collaborators \cite{Isoyama14,hami,Tiec:2015cxa}, in which expressions were derived through $O(\eta^2)$ for the (Bondi-like) energy and angular momentum of a circular binary with time-symmetric boundary conditions (so that spacetime admits a global helical symmetry). These expressions were given in terms of the so-called ``redshift'' variable \cite{Detweiler:2008ft} $z:=1/\hat u^t$, where $\hat u^{\alpha}$ is the four-velocity of the circular orbit, normalized not in the background metric $g_{\alpha\beta}$ but rather in the smooth perturbed metric $g_{\alpha\beta}+h^R_{\alpha\beta}$. CB related Refs.\ \cite{Isoyama14,hami,Tiec:2015cxa}'s notions of energy and angular momentum to $\EADM$ and $\LADM$, and thereby obtained an expression for  $\delta\LADM_c^{\rm cons}$ in terms of the redshift $z$. 

The expression they derived is remarkably simple:
\begin{equation}\label{Z}
\delta\LADM_c^{\rm cons}(\EADM) = -\lim_{\eps\to 0} \delta z(\EADM;\eps),
\end{equation}
where $\delta z$ is the first-order GSF correction to $z$, defined through the expansion
\begin{equation}
z(\EADM;\eps)=z^{(0)}(\EADM;\eps)+\delta z(\EADM;\eps) +O(\eta^2).
\end{equation}
Here $\EADM$ is used to parametrize the circular orbit, $z^{(0)}$ is the geodesic limit of $z$, and the $O(\eta)$ GSF correction $\delta z$ is defined with a fixed $\EADM$ (and a fixed $\eps$).\footnote{Beware that, in related literate, the GSF correction to $z$ is often defined with a fixed orbital frequency $\Omega$, not with a fixed $\EADM$ as here.} We note that, at our order of approximation, it is permissible to evaluate Eq.\ (\ref{Z}) using a sequence of {\em geodesic} circular orbits, replacing the ADM-related quantity $\EADM$ with the geodesic specific energy $E=-u_{t}$.  

The relation (\ref{Z}) is valid for all $\EADM>\EADM_{\rm isco}$ and can be used to derive $\delta\LADM_c^{\rm cons}$ for any unstable circular orbit, without the restriction $\EADM\geq 1$. Furthermore, the evaluation of $\delta\LADM_c^{\rm cons}$ using (\ref{Z}) requires only circular-orbit GSF data, readily computable using existing codes. In fact, using Eqs.\ (\ref{Z}), (\ref{calE+}) and (\ref{calW+}) it is evidently possible to evaluate both censorship condition (\ref{OS_weak}) (for weak fine-tuning) and (\ref{OS_strong}) (for strong fine-tuning) using circular-orbit information only. Our calculation in the next two sections takes a full advantage of that fact.

\section{Evaluation of censorship condition}\label{Sec:conservative}

In this section we evaluate the censorship condition (\ref{OS_weak}), which ignores the possibility of strong fine-tuning. The latter will be considered in Sec.\ \ref{Sec:dissipative}. At our order of approximation, we may replace the ADM-related quantity $\EADM$ in Eq.\ (\ref{OS_weak}) with the geodesic specific energy $E$, and regard (\ref{OS_weak}) as a condition on the family of (unstable) circular {\em geodesic} orbits, evaluated in the limit $\eps\to 0$. If the inequality can be shown to hold for all $E> E_{\rm isco}=\frac{1}{\sqrt{3}}$, then overspinning is ruled out for all orbits (except, possibly, strongly fine-tuned ones).

The evaluation of $(\ref{OS_weak})$ requires only the function $\delta\breve\LADM_c^{\rm cons}(E)=\delta \LADM_c^{\rm cons}(E)/\eta$, and we shall use Eq.\ (\ref{Z}) to calculate it. As will be shown, the evaluation of the redshift correction $\delta z$ in Eq.\ (\ref{Z}) becomes particularly simple in the limit $\eps\to 0$, to the effect that the essential part of the calculation can be done analytically. The only numerical input we shall require is a verification that a certain perturbative quantity has a finite limit $\eps\to 0$; the precise numerical value of that limit will not be important to us.  Below we first present the analytical part of the calculation, and in subsection (\ref{subsec:NumResultsZ}) we discuss the numerical input.

\subsection{Analytical considerations}

Isoyama {\it et al.}~\cite{Isoyama14} show that the first-order GSF correction to the redshift $z$ can be obtained via 
\begin{equation}\label{deltaz}
\delta z=-z^{(0)}H^R\quad \text{where} \quad H^R:=\frac{1}{2}h_{\alpha\beta}^{R,{\rm sym}} u^{\alpha}u^{\beta}.
\end{equation}
Here $z^{(0)}$ (recall) is the geodesic limit of $z$ (taken with fixed $\EADM,\eps$), $u^{\alpha}$ is the four-velocity of the corresponding circular geodesic, and $h_{\alpha\beta}^{R,{\rm sym}}$ is the ``time-symmetric'' part of the $R$ field evaluated on that geodesic. More precisely, $h_{\alpha\beta}^{R,{\rm sym}}$ is a certain regular piece not of the physical, retarded metric perturbation, but of a time-symmetrized, ``half-retarded plus half-advanced'' perturbation that is responsible for the ``conservative'' part of the dynamics. 

The evaluation of $\delta\LADM_c^{\rm cons}$ via Eq.\ (\ref{Z}) requires taking the limit $\eps\to 0$ of $\delta z$ with fixed $E$. To evaluate the limit of the factor $z^{(0)}$, start with the general expression \cite{Isoyama14}
\begin{equation}\label{z0}
z^{(0)}=(1-a\Omega)^{1/2}\left[1+a\Omega-3(M\Omega)^{2/3}(1-a\Omega)^{1/3}\right]^{1/2},
\end{equation}
in which $\Omega:=d\phi/dt=[a+M(R/M)^{3/2}]^{-1}$ is the orbital angular velocity, with $R$ being the Boyer-Linquist radius of the orbit.  The latter admits the small-$\eps$, fixed-$E$ expansion \cite{CB}
\begin{equation}
\label{rhopar}
R/M= 1+\eps \rho_1(E) +\eps^2 \rho_2(E) +O(\eps^3) ,  
\end{equation}
where the first two coefficients, needed below, read
\begin{eqnarray}
\label{rho12}
\rho_1&=& 4E(6E^2-2)^{-1/2}, 
\\
\rho_2&=& 2(2E^4-E^2+1)(3E^2-1)^{-2}.
\end{eqnarray}
Combined with $a/M=1-\eps^2$, this gives
\begin{equation}\label{OmegaExpansion}
M \Omega=\frac{1}{2}-\frac{1}{4}b(E)\,\eps +O(\eps^2),
\end{equation}
where $b$ was defined in (\ref{b}). Plugging (\ref{OmegaExpansion}) into Eq.\ (\ref{z0}) yields, in turn,
\begin{equation}\label{z0expansion}
z^{(0)}(E;\eps)=\frac{\eps}{\sqrt{6E^2-2}}+O(\eps^2).
\end{equation}
Finally, using (\ref{Z}) with (\ref{deltaz}) and (\ref{z0expansion}), we obtain
\begin{equation}\label{Z2}
\delta\LADM_c^{\rm cons}(E) =(6E^2-2)^{-1/2}\lim_{\eps\to 0}\left[\eps H^R(E;\eps)\right].
\end{equation}
Note that, for $\delta\LADM_c^{\rm cons}$ to be finite and generally nonzero (as expected) requires $H^R$ to blow up like $1/\eps$ for $\eps\to 0$.

Our calculation of $H^R(E;\eps)$ will be based on the strategy and numerical codes developed in Refs.\ \cite{Shah:2012gu,vandeMeent:2015lxa,meent:2015a}.
$H^R(E;\eps)$ is expressed as a sum of two contribution:
\begin{equation}\label{recons+compl}
H^R = H^R_{\rm recons} + H^R_{\rm compl}.
\end{equation}
The ``reconstructed'' part $H^R_{\rm recons}$ is obtained numerically, starting from frequency-domain solutions of Teukolsky's equation with a circular-geodesic source and retarded boundary conditions, following through a reconstruction of the multipole modes of the metric perturbation (in a suitable ``half-string'' radiation gauge \cite{PMB}), and finally applying a suitable form of mode-sum regularization \cite{Barack:1999wf} to extract the $R$ part of the perturbation. 
 (A more detailed description will be given in the next subsection.) In our case of a circular-orbit source, the double contraction of $h_{\alpha\beta}^{R}$ with $u^\alpha$ [recall Eq.\ (\ref{deltaz})] automatically picks out the time-symmetric part of $h_{\alpha\beta}^{R}$, as desired.

The second contribution to $H^R$ in Eq.\ (\ref{recons+compl}) is the ``completion'' piece $H^R_{\rm compl}$, which (by definition) arises from any part of the metric perturbation that is not captured by the reconstruction procedure. In our problem, this piece corresponds simply to mass and angular-momentum perturbations of the background Kerr geometry (plus pure-gauge perturbations) \cite{waldrec,Shah:2012gu}. In the vacuum region $r>R$ outside the particle's orbit, these stationary perturbations can be written analytically, in a ``Boyer-Lindquist'' gauge, as \cite{Shah:2012gu} 
\begin{equation}\label{dMdJ}
h_{\alpha\beta}^{(\delta M)}=\mu E \frac{\partial g_{\alpha\beta}}{\partial M},
\quad\quad
h_{\alpha\beta}^{(\delta J)}=\mu L \frac{\partial g_{\alpha\beta}}{\partial J},
\end{equation}
where $g_{\alpha\beta}=g_{\alpha\beta}(x^{\mu};M,J)$ is the Kerr background metric, $\partial_M$ is taken with fixed $J:=Ma$, $\partial_J$ is taken with fixed $M$, and both derivatives are taken with fixed Boyer-Lindquist coordinates $x^\mu$. Our particular regularization procedure (see below) does not require the completion piece for $r<R$. The quantity $H^R_{\rm compl}$ is given by
\begin{equation}\label{Hcompl}
H^R_{\rm compl}=
\frac{1}{2}u^{\alpha}u^{\beta}\left(h_{\alpha\beta}^{(\delta M)}+h_{\alpha\beta}^{(\delta J)}\right),
\end{equation}
where the perturbations are evaluated in the sided limit $r\to R^+$ (with $\theta=\pi/2$).

Let us denote by $\delta\LADM_{c,{\rm recons}}^{\rm cons}$ and $\delta\LADM_{c,{\rm compl}}^{\rm cons}$ the contributions to $\delta\LADM_{c}^{\rm cons}$ from $H^R_{\rm recons}$ and $H^R_{\rm compl}$, respectively,
and proceed to obtain $\delta\LADM_{c,{\rm compl}}^{\rm cons}$ analytically. First use Eq.\ (\ref{Hcompl}) with Eq.\ (\ref{dMdJ}) and with $u^{\alpha}=g^{\alpha\beta}u_{\beta}$, where $u_{\beta}=\{-E,0,0,L\}$ (in Boyer-Lindquist coordinates). This gives $H^R_{\rm compl}$ in terms of the circular-orbit radius $R$, its energy $E$ and its angular momentum $L$. Then substitute the fixed-$E$ expansions (\ref{rhopar}) for $R$, and \cite{CB}
\begin{equation}\label{LofE}
L/M=2E+(6E^2-2)^{1/2}\eps +O(\eps^2)
\end{equation}
for $L$, along with $a=1-\eps^2$. Finally, expand the result in $\eps$ at fixed $E$. The outcome is 
\begin{equation}
H^R_{\rm compl}=\frac{\eta}{2\eps}(1-E^2)(6E^2-2)^{1/2} + O(\eps^0).
\end{equation}
Notice this is an $O(\eps^{-1})$ quantity, so, recalling Eq.\ (\ref{Z2}), it gives a finite contribution to $\delta\breve\LADM_c^{\rm cons}$. We find
\begin{equation}\label{Z3}
\delta\breve\LADM_{c,{\rm compl}}^{\rm cons} =\frac{1}{2}(1-E^2), 
\end{equation}
where $\delta\breve\LADM_{c,{\rm compl}}^{\rm cons}:=\eta^{-1}\delta\LADM_{c,{\rm compl}}^{\rm cons}$.

Remarkably, it follows that the completion contribution, on its own, {\em precisely saturates} the censorship condition (\ref{OS_weak}).  In the next subsection we will demonstrate numerically that the reconstructed part, $H^{R}_{\rm recons}(E;\eps)$, has a {\it finite} (non-divergent) fixed-$E$ limit $\eps\to 0$. This will imply
\begin{equation}\label{Hrecons}
\delta\LADM_{c,{\rm recons}}^{\rm cons}=0,
\end{equation}
and therefore
\begin{equation}\label{saturation}
\delta\breve\LADM_{c}^{\rm cons} =\frac{1}{2}(1-E^2). 
\end{equation}
The censorship condition (\ref{OS_weak}) is precisely saturated. 
This result and its implications will be discussed in Sec.\ \ref{Sec:conclusions}.

\subsection{Numerical input} \label{subsec:NumResultsZ}

To validate Eq.\ (\ref{Hrecons}), we will demonstrate numerically that the limit
\begin{equation}\label{hatHR}
\lim_{\eps\to 0}H^R_{\rm recons}(E;\eps) =:  \hat H^R(E)
\end{equation}
(taken with fixed $E$) exists and yields a finite value. It may be possible to establish this mathematically through analysis of the reconstructed solutions in the near-extremal near-horizon approximations (perhaps modelled upon the method of Ref.\ \cite{Gralla:2015rpa}). Here we content ourselves with a numerical calculation, which, we find, already illustrates the finiteness of $\hat H^R(E)$ rather convincingly. 

We have performed two independent numerical calculations, using two different (albeit related) methods, to be described below. One of the methods performs best at $\eps$ values that are not too small, and the other does best for $\eps$ values that are not too large. The combination of the two methods thus allowed us access to a range of $\eps$ values wide enough to enable taking the limit $\eps\to 0$ accurately. The agreement we found between the two sets of results in an  overlapping domain of intermediate $\eps$ values provides reassurance.    

Our first calculation is based on method and code developed by one of us (AGS, with collaborators) in Refs.\ \cite{Shah:2010bi,Shah:2012gu}, with input from Ref.\ \cite{vandeMeent:2015lxa}. In this method, we first numerically integrate the sourced Sasaki-Nakamura equation in the frequency domain, with ``retarded'' boundary conditions at infinity and on the event horizon, to obtain the modes of the Weyl scalar $\psi_0$ associated with the metric perturbation produced by the particle. We then derive an appropriate Hertz potential (this is done algebraically in terms of $\psi_0$), from which the modes of the metric perturbation are reconstructed by applying a certain second-order differential operator \cite{Keidl:2010pm}. We use a version of the reconstruction procedure that yields the metric perturbation in a regular outgoing radiation gauge anywhere in the vacuum region $r>R$, where $R$ is the Boyer-Lindquist radius of the circular orbit. Finally, we apply a mode-sum regularization procedure to obtain $H^R_{\rm recons}$. The mode-sum variant we are using is the one developed in \cite{vandeMeent:2015lxa}, with the particle limit taken from $r\to R^+$. (Ref.\ \cite{vandeMeent:2015lxa} derived the regularization parameter values suitable for this one-sided version of the mode-sum method.) The code is implemented in \verb!C++! and uses double-precision arithmetic.

This has been a first implementation of the code in the near-extremal regime, $\eps\ll 1$. Certain technical subtleties arise in this regime, as recently reviewed in Sec.\ V of \cite{Gralla:2015rpa}. We have found that such subtleties were rather easily controlled for $\eps\gtrsim 10^{-4}$. We needed only ensure that inner boundary conditions were placed sufficiently close to the horizon and determined to sufficient accuracy. We have not attempted to improve the performance of the code at lower values of $\eps$ (e.g., using the techniques described in Ref.\ \cite{Gralla:2015rpa}), but instead resorted to our second method, to be described next, whose performance actually improves near extremality. 

Our second calculation is based on a code developed by one of us (MvdM) in \cite{meent:2015a}, which follows an approach by Fujita \cite{Fujita:2004rb}, itself based on the semi-analytical formalism of Mano, Suzuki, and Takasugi (MST) \cite{Mano:1996vt,Mano:1996gn}. In this approach, the Weyl scalar---$\psi_4$ in our particular implementation---is obtained semi-analytically rather than numerically. ``Semi'' here refers  to an element of the calculation in which a certain continued-fraction equation is solved numerically. The reconstruction and mode-sum procedures are essentially as in the first method, but they are implemented using an independent code. The entire calculation is performed using {\it Mathematica} with arbitrary-precision arithmetic. 

In the MST-based calculation, working near extremality is computationally advantageous. This is due to the improved convergence properties of the MST formalism for cicular orbits with $a\sim 1$ and $\Omega\sim 1/2$, highlighted in \cite{meent:2015a}. In this domain, the series of special functions featuring in MST's solutions for $\psi_4$ converges faster. Furthermore, the aforementioned continued-fraction equation is both faster convergent and more easily solvable (using the analytically-known extremal solution as an initial guess). As a result, the method is particularly efficient for studying the $\eps\to 0$ limit. For our purpose, it was sufficient to apply it in the range $10^{-8}\leq \eps\leq 10^{-4}$. Above $\eps\sim 10^{-4}$, the analytically-known extremal solution no longer provides an accurate enough initial guess to guarantee finding the solution of the continued-fraction equation for all frequency modes in the spectrum, making our implementation of the MST formalism unreliable. 

Our calculation of $\hat H^R(E)$ proceeded as follows. We considered a dense sample of $E$ values in the range $E_{\rm isco}<E<2$. For each value of $E$ in the sample we obtained a dataset $H^R_{\rm recons}(E,\eps)$, where $\eps$ is sampled (roughly) uniformly in $\log\eps$ between $\eps=10^{-1}$ and $\eps=10^{-8}$. We switched from our fully numerical method to our MST-based method at around $\eps=10^{-4}$.  $\hat H^R(E)$ was then obtained via extrapolation of each of the fixed-$E$ datasets to $\eps=0$.  

For each pair $\{E,\eps\}$ in our sample, we directly computed the first 70 multipoles ($l$-modes) of the metric perturbation, for use as input in the mode-sum formula. The remaining large-$l$ tail of modes was approximated by fitting an inverse-power-law model, as detailed in \cite{Shah:2012gu}. At high values of $E$, the $l$-mode distribution becomes skewed towards larger $l$ values, due to what may be interpreted as a beaming effect. A similar behavior near the light-ring of a Schwarzschild black hole was discussed by Akcay {\it et al.} \cite{Akcay}, who pointed out that the implementation of the mode-sum technique can become problematic in that case, because the standard inverse-power-law tail may fail to manifest itself until $l$ values larger than one can feasibly calculate. This effect restricted our calculation here to $E$ values that are not too large---in practice, to $E\lesssim 2$. However, that should suffice for our purpose here, which is simply to determine the $\eps$-scaling of $H^R_{\rm recons}$ in the limit $\eps\to 0$ at fixed $E$: it is perfectly reasonable to assume that the $\eps$-scaling at any fixed $E>2$ would be the same as it is for lower $E$.

\begin{figure}[htb]
\includegraphics[width=\columnwidth]{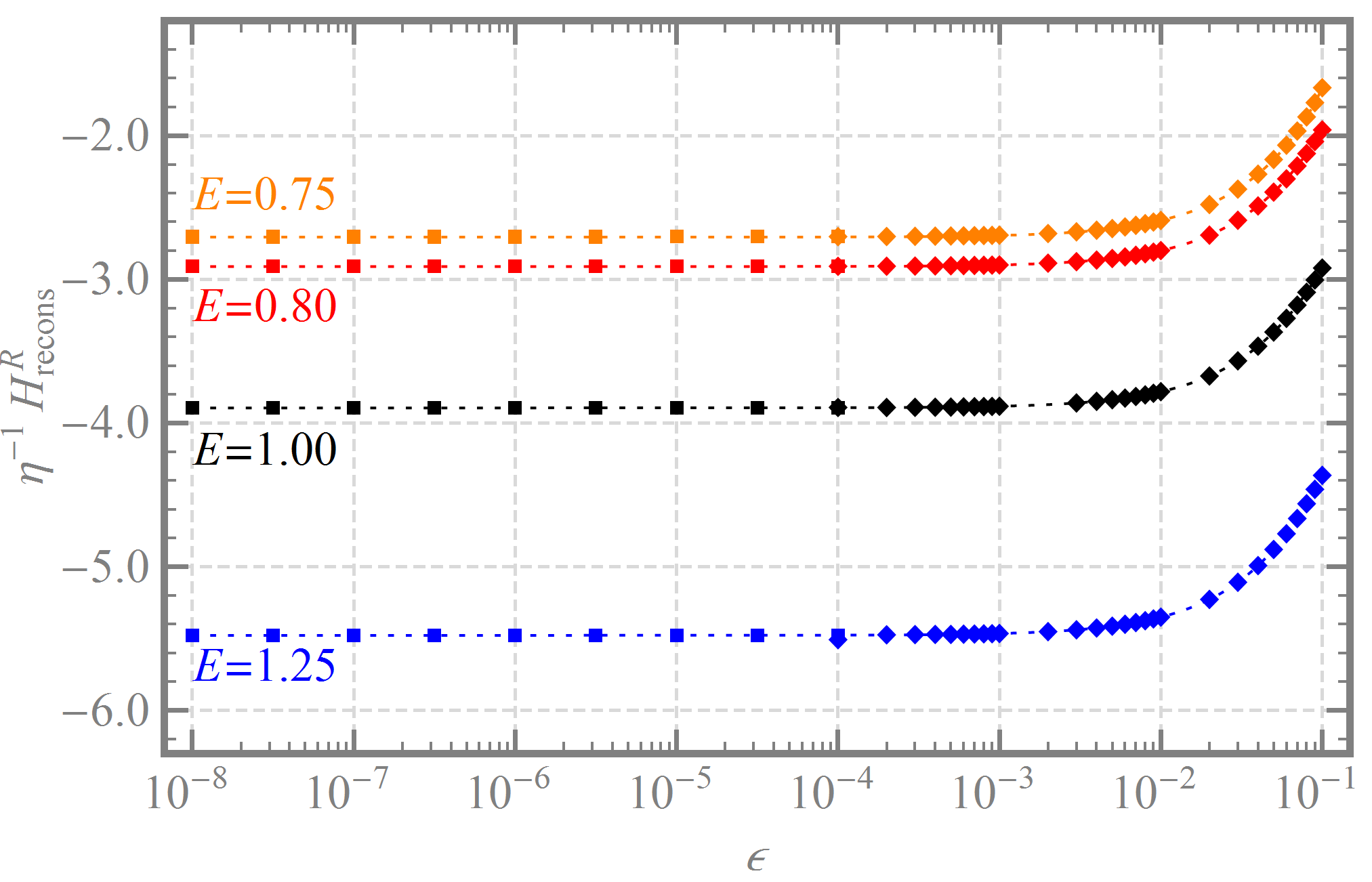}
\caption{$H^R_{\rm recons}$ as a function of $\eps$, for a sample of $E$ values. Data points for $\eps\geq 10^{-4}$ (diamonds) are from our fully numerical computation, while points for $\eps\leq 10^{-4}$ (squares) are from our semi-analytical, MST-based method (there is an overlapping data point at $\eps=10^{-4}$). Error bars are in all cases too small to be resolved in this figure. Curves (dotted line) are quartic polynomial fits.  At each fixed $E$, $H^R_{\rm recons}(E;\eps)$ appears to approach a constant value in the extremal limit $\eps\to 0$. 
}
\label{fig:Hvseps}
\end{figure}

\begin{figure}[htb]
\includegraphics[width=\columnwidth]{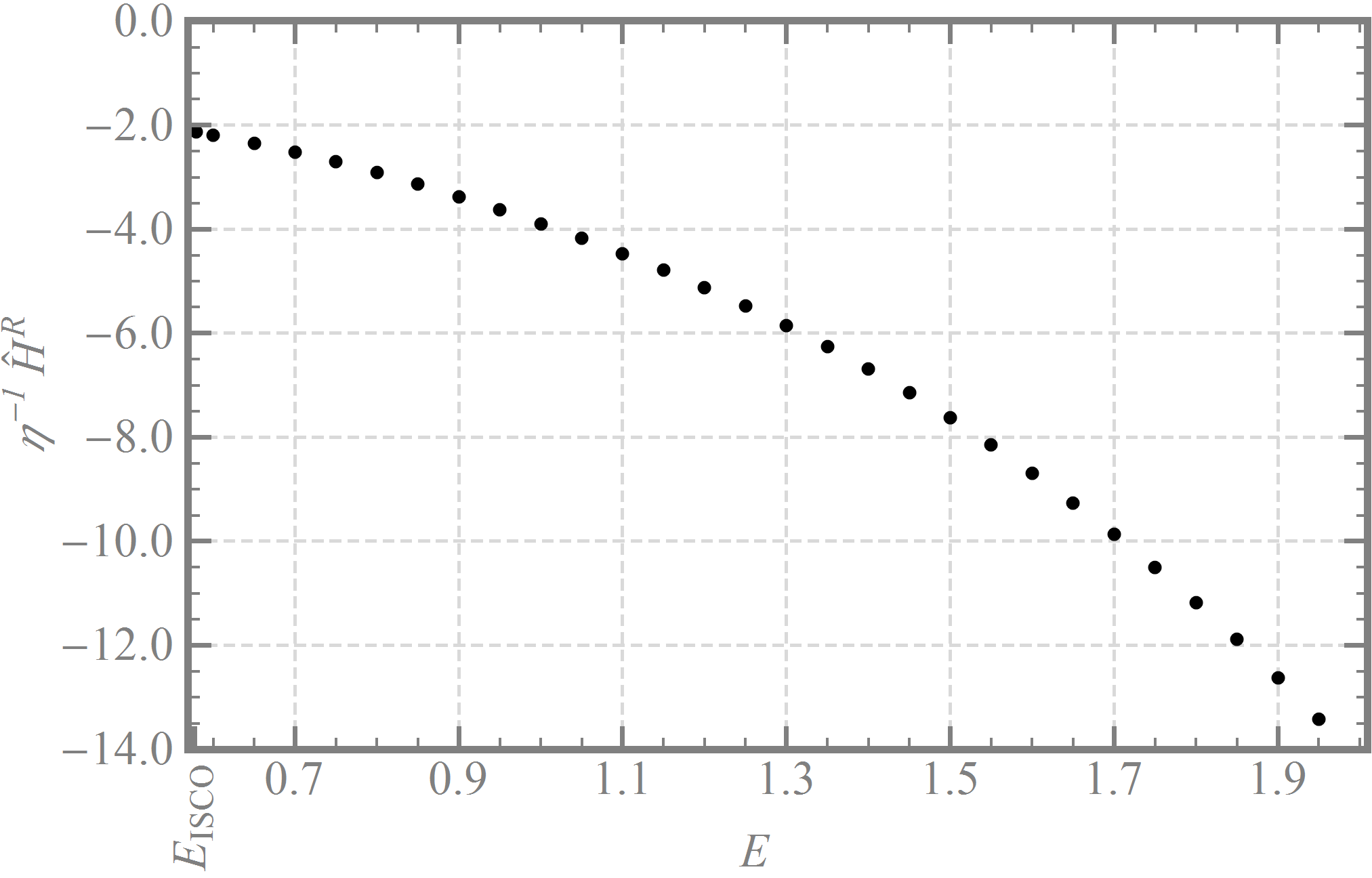}
\caption{The function $\hat H^R(E)$, obtained by extrapolating our numerical data for $H^R_{\rm recons}$ to $\eps\to 0$ at each fixed $E$. 
The actual value of $\hat H^R(E)$ is not needed in our analysis, only the fact that it is finite for each finite $E$. 
}
\label{fig:HvsE}
\end{figure}

Figure \ref{fig:Hvseps} shows $H^R_{\rm recons}(E;\eps)$ as a function of $\eps$ for a few $E$ values within our sample. It is evident that $H^R_{\rm recons}(E;\eps)$ approaches a finite limit as $\eps\to 0$. Figure \ref{fig:HvsE} displays the extrapolated values $\hat H^R$ as a function of $E$. We remind that the details of the function $\hat H^R(E)$ are unimportant to us; we needed only establish here that $\hat H^R$ is finite for any finite $E$. 

\section{Effect of strong fine-tuning}\label{Sec:dissipative}

We have shown that, within our first-order GSF approximation, any weakly fine-tuned capture produces a precisely extremal geometry. Can strong fine-tuning push the system beyond extremality? To answer that question we need to evaluate the condition (\ref{OS_strong}). Any choice of $\{E_i,E_f\}$ (with $E_{\rm isco}\leq E_f<E_i$) violating that condition would imply that overspinning is possible via strong fine-tuning. If, on the other hand, we can show that (\ref{OS_strong}) applies for any $\{E_i,E_f\}$, then censorship holds even allowing for strong fine-tuning. 

The evaluation of (\ref{OS_strong}) requires the angular-momentum shift $\delta\breve\LADM_c^{\rm cons}(E_i)$ and the flux ratio ${\cal R}(E)$. For the former we use our result (\ref{saturation}). For the latter we will perform a numerical calculation, to be presented in subsection \ref{subsec:NumResults} below. However, much of what we need to know about ${\cal R}$ can be deduced from simple analytic considerations, to be presented first. We will show that it is sufficient to require that ${\cal R}$ is bounded from below by $-1/3$ over the range $E_{\rm isco}\leq E<\frac{2}{\sqrt{3}}$ in order to guarantee that the censorship condition (\ref{OS_strong}) always holds. Our actual calculation will later show that $\cal R$ lies very comfortably above that bound. 

\subsection{Analytical considerations}

\subsubsection{Superradiant domain}

First, we consider the sign of ${\cal R}(E)$. We recall that this function, defined for circular equatorial geodesics, is the ratio of energy flux through the horizon to the energy flux at infinity. A gravitational-wave mode of the form $\sim e^{-i\omega t}e^{im\varphi}$ is known to be superradiant if and only if $0<\omega <m\Omega_H$, where $\Omega_H=a/(2Mr_+)$ is the horizon's angular velocity, with $r_+=M+(M^2-a^2)^{1/2}$. For circular equatorial orbits the gravitational-wave spectrum is simple: $\omega=m\Omega$, where $\Omega$ is the orbital angular velocity. Thus, all modes of the radiation are superradiant for $\Omega<\Omega_H$, giving ${\cal R}<0$ in that case. For $\Omega>\Omega_H$ all modes are instead non-superradiant, giving ${\cal R}>0$.

Let us now specialize to a near-extremal geometry, and reexpress the above in terms of a condition on the specific energy $E$ of the circular geodesic. For $\eps\ll 1$ we find $M\Omega_H=\frac{1}{2}-\frac{1}{\sqrt{2}}\eps+O(\eps^2)$. Combining this with Eq.\ (\ref{OmegaExpansion}) translates the superradiance condition $\Omega<\Omega_H$ to $E(6E^2-2)^{-1/2}>\sqrt{2}/3$ (at leading order in $\eps$), leading to 
\begin{equation}
E<\frac{2}{\sqrt{3}} =: E_{\rm sr}
\end{equation}
as the superradiant domain in the extremal limit. Thus
\begin{eqnarray}
{\cal R} &<& 0\quad \text{for}\quad E_{\rm isco}\leq E<E_{\rm sr}, 
\\
{\cal R} &>& 0\quad \text{for}\quad E>E_{\rm sr}. \label{noSR}
\end{eqnarray}
This will be confirmed numerically in subsection \ref{subsec:NumResults}.

\subsubsection{Sufficient lower bound for ${\cal R}(E)$}

We now show that the censorship condition 
(\ref{OS_strong}) is satisfied for all $E_i>E_f\geq \frac{1}{\sqrt{3}}$ if $\calR(E)$ is bounded from below by $-1/3$. The condition involves $\tilde\phi$ and $\tilde\psi$, given in Eqs.\ (\ref{tildephi}) and (\ref{tildepsi}), respectively, where in the latter we now substitute for $\delta\breve\LADM_c^{\rm cons}$ from Eq.\ (\ref{saturation}). We do not know the sign of $\tilde\phi$ (for given $E_i,E_f$) a priori, so we proceed by considering the two options $\tilde\phi\geq 0$ and $\tilde\phi<0$ in turn.

{\it Case $\tilde\phi\geq 0$}:--- The censorship condition (\ref{OS_strong}) becomes 
\begin{equation}\label{OS_strong1}
\tilde\psi=-\frac{1}{2}(1-E_i^2)+(E_i-\breve{\cal E}^+_{\rm (qc)})^2 \geq 0.
\end{equation}
This is trivially satisfied for $E_i\geq 1$, so it remains to consider $E_i<1$, in which case the condition becomes 
\begin{equation}\label{OS_strong2}
\breve{\cal E}^+_{\rm (qc)}\leq E_i-\sqrt{(1-E_i^2)/2}=:\nu_1(E_i).
\end{equation}
Recalling (\ref{calE+}), we may bound $\breve{\cal E}^+_{\rm (qc)}$ from above using
\begin{equation}\label{OS_strong3}
\breve{\cal E}^+_{\rm (qc)}
\leq \int_{\frac{1}{\sqrt{3}}}^{E_i} \frac{dE}{1+{\cal R}(E)} 
\leq \frac{E_i-\frac{1}{\sqrt{3}}}{1+{\cal R}_m} =:\nu_2(E_i),
\end{equation}
where in the first inequality we used the positivity of the integrand together with $E_f\geq\frac{1}{\sqrt{3}}$, and in the second inequality we assumed $\cal R$ is bounded from below by some number ${\cal R}_m(>-1)$. 
Since $\nu_1=\nu_2(=0)$ at $E_i=\frac{1}{\sqrt{3}}$, establishing the inequality in (\ref{OS_strong2}) requires only showing that $\nu_1'(E_i)\geq \nu_2'(E_i)=(1+{\cal R}_m)^{-1}$ for all $\frac{1}{\sqrt{3}}<E_i<1$. But the minimal value of $\nu_1'$ over this domain is $3/2$, so the condition becomes $(1+{\cal R}_m)^{-1}\leq \frac{3}{2}$, or ${\cal R}_m\geq -\frac{1}{3}$.
We have thereby shown that the censorship condition (\ref{OS_strong1}) holds for any $E_i>E_f\geq\frac{1}{\sqrt{3}}$ with $\tilde\phi(E_i,E_f)\geq 0$, under the sole assumption
\begin{equation}\label{calRm}
{\cal R}(E)\geq -\frac{1}{3}. 
\end{equation}

{\it Case $\tilde\phi< 0$}:--- The censorship condition (\ref{OS_strong}) becomes $\tilde\psi\geq \tilde\phi^2/4$, or, explicitly,
\begin{eqnarray}
0&\leq& -\frac{1}{4}\breve{\cal W}^+_{\rm (qc)}\left(\breve{\cal W}^+_{\rm (qc)}-2\phi(E_i)\right)
+\breve{\cal E}^+_{\rm (qc)}\left(\breve{\cal E}^+_{\rm (qc)}-2E_i\right)
\nonumber\\
&=:&\Delta(E_i,E_f),
\end{eqnarray}
where $\phi(E_i)=-(6E_i^2-2)^{1/2}$. Since $\breve{\cal W}^+_{\rm (qc)}=0=\breve{\cal E}^+_{\rm (qc)}$ for $E_i=E_f$, we have $\Delta(E,E)=0$ for all $E\geq\frac{1}{\sqrt{3}}$. Thus, to establish $\Delta\geq 0$ it suffices to show $\partial\Delta(E_i,E_f)/\partial E_i\geq 0$ for all $E_i\geq E_f\geq\frac{1}{\sqrt{3}}$. 

With the aid of Eqs.\ (\ref{calE+}) and (\ref{calW+}), we find
\begin{equation}\label{dDelta}
[1+{\cal R}(E_i)]\frac{\partial\Delta}{\partial E_i}=
E_i + {\cal R}(E_i)\left[\frac{3E_i \breve{\cal W}^+_{\rm (qc)}}{\phi(E_i)} - 2\breve{\cal E}^+_{\rm (qc)}\right].
\end{equation}
Consider the cases ${\cal R}(E_i)\leq 0$ and ${\cal R}(E_i)> 0$ separately. For ${\cal R}(E_i)\leq 0$, we use $\breve{\cal W}^+_{\rm (qc)}>\phi(E_i)$ (following from $\tilde\phi< 0$) to bound the right-hand side of (\ref{dDelta}) from below by $E_i[1+3{\cal R}(E_i)]-2{\cal R}(E_i)\breve{\cal E}^+_{\rm (qc)}$. Since the last term here is non-negative, it is sufficient to require ${\cal R}(E_i)\geq-\frac{1}{3}$ in order to guarantee $\partial\Delta/\partial E_i> 0$ and hence $\Delta(E_i,E_f)\geq 0$. If, instead, ${\cal R}(E_i)> 0$, one can first use $- 2\breve{\cal E}^+_{\rm (qc)}>\breve{\cal W}^+_{\rm (qc)}$ [which follows from Eqs.\ (\ref{calE+}) and (\ref{calW+}), noting $b(E)>2$], then again $\breve{\cal W}^+_{\rm (qc)}>\phi(E_i)$, to bound the right-hand side of (\ref{dDelta}) from below by $E_i+{\cal R}(E_i)\left[\phi(E_i)+3E_i\right]$. This is non-negative for all $E_i\geq \frac{1}{\sqrt{3}}$ if and only if ${\cal R}(E_i)\geq -\frac{1}{3}$. Thus, the condition (\ref{calRm}) always implies $\Delta\geq 0$ and, in turn, that the censorship condition (\ref{OS_strong}) holds also for $\tilde\phi<0$.

We conclude that it is sufficient to show that the flux ratio $\cal R$ is bounded from below by $-\frac{1}{3}$ in order to guarantee that the censorship condition (\ref{OS_strong}) is always satisfied. In fact, recalling (\ref{noSR}), we see that it is sufficient to obtain such a bound for $\cal R$ on the restricted superradiant domain $E_{\rm isco}\leq E<E_{\rm sr}$. Our numerical calculation, to be presented below, shows that $\cal R$ is comfortably bounded above the value of $-1/3$.

\subsection{Numerical input} \label{subsec:NumResults}

To compute the flux ratio $\cal R$ we used our MST-based method described above, as implemented in \cite{meent:2015a}. The gravitational-wave energy fluxes to infinity and down the horizon are obtained directly from the semi-analytical solutions for $\psi_4$, with no need to reconstruct the metric perturbation. Thanks to the improved convergence properties (already mentioned above) of the MST formalism at $\eps\ll 1$ and $\Omega\sim 1/2$, we can obtain the energy fluxes 
to essentially any accuracy we desire using arbitrary-precision computer algebra.

\begin{figure}[htb]
\includegraphics[width=\columnwidth]{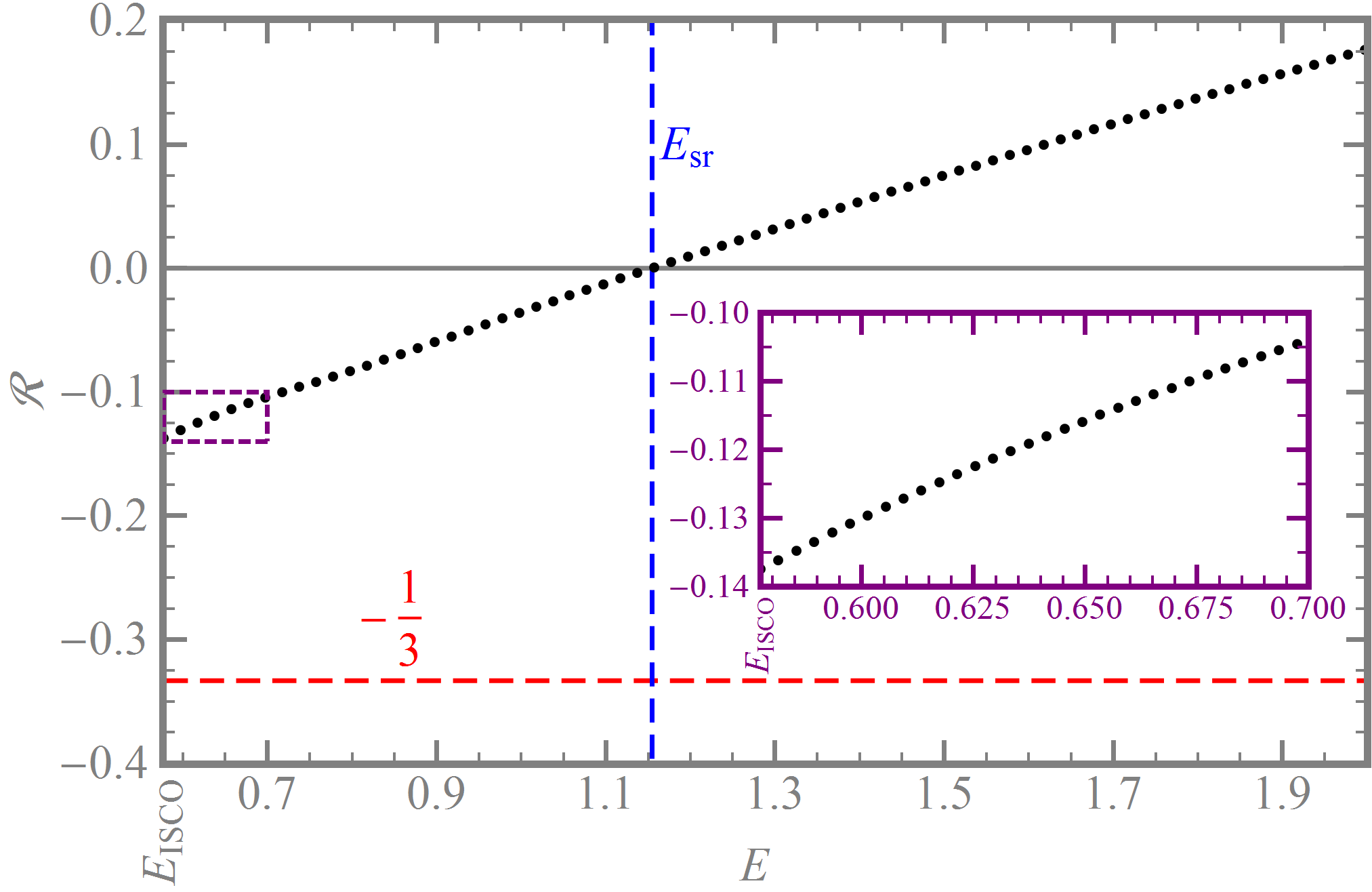}
\caption{Numerical values for the flux ratio $\cal R$, as a function of specific energy $E$, for unstable circular equatorial geodesic orbits in the extremal Kerr limit. Each dot represents a numerical measurement, with error bars being much smaller than the size of the dots. Orbits with $E<E_{\rm sr}=2/\sqrt{3}$ are superradiant, with ${\cal R}<0$. The inset expands the area around the minimum at $E_{\rm isco}=1/\sqrt{3}$. We find a minimum value of  ${\cal R}=-0.13744\pm 3\cdot 10^{-5}$, safely above what is required for the censorship condition \eqref{OS_strong} to hold (dashed red line).
}
\label{fig:Rplot}
\end{figure}

To determine $\cal R$ for a given $E$, we calculated the ratio between the flux down the horizon and the flux to infinity for a sequence of fixed-$E$ orbits with $\epsilon$ values that decrease to $10^{-8}$ in exponential steps. The value of ${\cal R}(E)$ was then found by extrapolating to $\epsilon=0$. The results are presented in Fig.\ \ref{fig:Rplot}. As expected, ${\cal R}(E)$ is negative only in the range $E_{\rm isco}\leq E<E_{\rm sr}$. The minimum of $\cal R(E)$ appears to be attained at $E_{\rm isco}$ with a value of $-0.13744\pm 3\cdot 10^{-5}$. This is comfortably above the value of $-1/3$ required to assure that the censorship condition \eqref{OS_strong} is satisfied.

\section{Conclusions and discussion}\label{Sec:conclusions}

We studied the scenario in which a mass particle is thrown into a nearly-extremal Kerr black hole on an equatorial trajectory, working consistently in the first-order self-force approximation, i.e.~taking into account all finite-$\eta$ effects (radiative and other) to one order in $\eta$ beyond the geodesic approximation. To describe the fate of the post-capture geometry we followed a strategy set out by CB in \cite{CB}, according to which it suffices to consider two types of captured orbits near the capture--scatter separatrix: (i) weakly fine-tuned orbits, which execute $O(\ln\eta)$ quasi-circular revolutions below the ISCO prior to falling into the black hole, radiating $O(\eta^2\ln\eta)$ of gravitational-wave energy in the process; and (ii) strongly fine-tuned orbits, which execute $O(\eta^{-1})$ revolutions and emit $O(\eta)$ of energy.

In Sec.\ \ref{Sec:conservative} we found that, within our first-order GSF approximation, any weakly fine-tuned capture leads to a precisely extremal geometry. This implies \cite{CB} that ``generic'' captures (ones that are not fine-tuned at all, with $\LADM-2\EADM$ negative and not small) produce {\em sub}-extremal geometries.  In Sec.\ \ref{Sec:dissipative} we further established that {\em strong fine-tuning promotes censorship}: all strongly fine-tuned captures produce subextremal geometries. Thus, one can at best reach extremality, using weakly fine-tuned orbits (any such orbit would do), but there is no way of overspinning the black hole. In summary: 
\begin{quote}
 Within the first-order self-force approximation (and excluding deeply bound orbits),
{\it equatorial captures generically result in a subextremal post-capture geometry. One can at best achieve extremality, through weak fine-tuning, but overspinning is not possible.}
\end{quote}
That overspinning appears to be possible in the geodesic approximation \cite{js} is simply an artefact of ignoring important GSF terms that appear {\it already at leading order} in the relevant overspinning conditions.

We note that the above conclusions were arrived at almost entirely via analytical considerations. We required only two pieces of numerical input, one confirming the boundedness of the extremal limit in Eq.\ (\ref{hatHR}), and another establishing the bound (\ref{calRm}) for the flux ratio.  Both numerical computations involve only circular geodesic orbits, and neither requires particularly high precision. The above general conclusions are entirely robust with respect to numerical error. 

However, it is important to remember that here we have been working strictly within the framework of the first-order GSF approximation, with no control whatsoever over high-order GSF corrections. Since the first-order analysis appeared to allow for an exact saturation of the overspinning condition, higher-order effects may qualitatively change the outcome. A second-order analysis may potentially yield any possible result: that the final geometry is always subextremal, or that overspinning is possible, or (once again) that the black hole can at most be brought to extremality. In that respect, {\it our first-order GSF analysis---just like the geodesic analysis of Ref.\ \cite{js}---does not provide a conclusive answer to the question of overspinning.} It is not clear if the question can be fully resolved at second-order or at any other finite order in perturbation theory.  This may be a disappointing conclusion, but it is an interesting one nonetheless. 


For whatever it is worth, let us return to discuss the consequences of our first-order analysis. We have found that overspinning is not possible, consistent with the conjecture of weak cosmic censorship. However, we have also founds that, through (weak) fine-tuning of the orbital parameters, one can drive the black hole to extremality. This is an intriguing possibility, because it appears to be in violation of Israel's ``third law'' of black hole dynamics \cite{Israel}. We have not studied in detail whether the conditions of the third law can be said to be met in full in our problem. Reference \cite{zimm} contains some discussion of this point, and proposes how any apparent violation of the third law in the overspinning problem might be reconciled. 

There are several ways in which our analysis may be improved and further tested. First, it would be desirable to repeat the calculation of the angular momentum shift $\delta\LADM_c^{\rm const}$ using a direct integration of the GSF, via Eq.\ (\ref{dLc}) (with $\tilde F_{\alpha}\to \tilde F_{\alpha}^{\rm cons}$). This would eliminate our reliance [in deriving Eq.\ (\ref{Z})] on the effective Hamiltonian formulation of Refs.\ \cite{Isoyama14,hami,Tiec:2015cxa}, which is axiomatic in nature. In fact,  an explicit demonstration of agreement between the direct formula (\ref{dLc}) and redshift formula (\ref{Z}) would constitute an important test of the Hamiltonian formulation. There is work in progress at Southampton to directly (numerically) evaluate $\delta\LADM_c^{\rm const}$ via Eq.\ (\ref{dLc}).

Second, it may be possible to replace some (or all) of the numerical input for our analysis with analytical arguments. Specifically, one may seek to establish analytically the boundedness of $\hat H^R$ in Eq.\ (\ref{hatHR}), and the lower bound (\ref{calRm}) for the flux ratio $\cal R$. Both may be achieved by extending the ISCO analysis of Ref.\  \cite{Gralla:2015rpa} to unstable circular orbits, and (in the case of $\hat H^R$) also from the Weyl scalar to the reconstructed metric perturbation. We leave this to future work.

To conclude, we reiterate our view that our work represents a first complete analysis of the overspinning/overcharging problem through first post-geodesic order in perturbation theory, for a particular capture scenario. Of course, we have only explored here a fraction of the space of interesting scenarios. Our analysis did not cover, for example, (i) very low energy configurations of deeply-bound orbits, (ii) non-equatorial orbits,  (iii) ultrarelativistic or null particles, or (iv) spinning and/or electrically charged particles on a Kerr-Newman background. These scenarios all deserve examination.



\section*{Acknowledgements}

We gratefully acknowledge support from the European Research Council under the European Union's Seventh Framework Programme FP7/2007-2013/ERC, Grant No.\ 304978.  LB acknowledges additional support from STFC through grant number PP/E001025/1.

\bibliography{biblio}

\end{document}